\documentstyle[twoside,fleqn,epsf]{article}
\makeatletter
\def\fileversion{v2.6}
\def\filedate{24 November 1993}

\newdimen\@bls                    
\@bls=\baselineskip               
\advance\@bls -1ex                
\newdimen\@eps                    %
\def\section{\@startsection{section}{1}{\z@}
  {1.5\@bls plus 0.5\@bls}{1\@bls}{\normalsize\bf}}
\def\subsection{\@startsection{subsection}{2}{\z@}
  {1\@bls plus 0.25\@bls}{\@eps}{\normalsize\bf}}
\def\subsubsection{\@startsection{subsubsection}{3}{\z@}
  {1\@bls plus 0.25\@bls}{\@eps}{\normalsize\bf}}
\def\paragraph{\@startsection{paragraph}{4}{\parindent}
  {1\@bls plus 0.25\@bls}{0.5em}{\normalsize\bf}}
\def\subparagraph{\@startsection{subparagraph}{4}{\parindent}
  {1\@bls plus 0.25\@bls}{0.5em}{\normalsize\bf}}

\def\@sect#1#2#3#4#5#6[#7]#8{\ifnum #2>\c@secnumdepth
  \def\@svsec{}\else
  \refstepcounter{#1}\edef\@svsec{\csname the#1\endcsname.\hskip0.5em}\fi
  \@tempskipa #5\relax
  \ifdim \@tempskipa>\z@
    \begingroup
      #6\relax
      \@hangfrom{\hskip #3\relax\@svsec}{\interlinepenalty \@M #8\par}%
    \endgroup
    \csname #1mark\endcsname{#7}\addcontentsline
      {toc}{#1}{\ifnum #2>\c@secnumdepth \else
        \protect\numberline{\csname the#1\endcsname}\fi #7}%
  \else
    \def\@svsechd{#6\hskip #3\@svsec #8\csname #1mark\endcsname
      {#7}\addcontentsline{toc}{#1}{\ifnum #2>\c@secnumdepth \else
        \protect\numberline{\csname the#1\endcsname}\fi #7}}%
  \fi \@xsect{#5}}

\long\def\@makefigurecaption#1#2{\vskip 10mm #1. #2\par}

\long\def\@maketablecaption#1#2{\hbox to \hsize{\parbox[t]{\hsize}
  {#1 \\ #2}}\vskip 0.3ex}

\def\fnum@figure{Figure \thefigure}
\def\figure{\let\@makecaption\@makefigurecaption \@float{figure}}
\@namedef{figure*}{\let\@makecaption\@makefigurecaption \@dblfloat{figure}}

\def\table{\let\@makecaption\@maketablecaption \@float{table}}
\@namedef{table*}{\let\@makecaption\@maketablecaption \@dblfloat{table}}

\textfloatsep=\floatsep           
\intextsep=\floatsep              

\long\def\@makefntext#1{\parindent 1em\noindent\hbox{${}^{\@thefnmark}$}#1}

\def\maketitle{\begingroup        
    \def\thefootnote{\fnsymbol{footnote}}%
    \newpage \global\@topnum\z@
    \@maketitle \@thanks
  \endgroup
  \let\maketitle\relax \let\@maketitle\relax
  \gdef\@thanks{}\let\thanks\relax
  \gdef\@address{}\gdef\@author{}\gdef\@title{}\let\address\relax}

\def\justify@on{\let\\=\@normalcr
  \leftskip\z@ \@rightskip\z@ \rightskip\@rightskip}

\newbox\fm@box                    

\def\@maketitle{
  \global\setbox\fm@box=\vbox\bgroup
    \vskip 8mm                    
    \raggedright                  
    \hyphenpenalty\@M             
    {\Large \@title \par}         
    \vskip\@bls                   
    {\normalsize                  
     \@author \par}               
    \vskip\@bls                   
    \@address                     
  \egroup
  \twocolumn[
    \unvbox\fm@box                
    \vskip\@bls                   
    \unvbox\abstract@box          
    \vskip 2pc]}                  

\newcounter{address}
\def\theaddress{\alph{address}}
\def\@makeadmark#1{\hbox{$^{\rm #1}$}}

\def\address#1{\addressmark\begingroup
  \xdef\@tempa{\theaddress}\let\\=\relax
  \def\protect{\noexpand\protect\noexpand}\xdef\@address{\@address
  \protect\addresstext{\@tempa}{#1}}\endgroup}
\def\@address{}

\def\addressmark{\stepcounter{address}%
  \xdef\@tempb{\theaddress}\@makeadmark{\@tempb}}

\def\addresstext#1#2{\leavevmode \begingroup
  \raggedright \hyphenpenalty\@M \@makeadmark{#1}#2\par \endgroup
  \vskip\@bls}

\newbox\abstract@box              

\def\abstract{%
  \global\setbox\abstract@box=\vbox\bgroup
  \small\rm
  \ignorespaces}
\def\endabstract{\par \egroup}

\def\thebibliography#1{\section*{REFERENCES}\list{\arabic{enumi}.}
  {\settowidth\labelwidth{#1.}\leftmargin=1.67em
   \labelsep\leftmargin \advance\labelsep-\labelwidth
   \itemsep\z@ \parsep\z@
   \usecounter{enumi}}\def\makelabel##1{\rlap{##1}\hss}%
   \def\newblock{\hskip 0.11em plus 0.33em minus -0.07em}
   \sloppy \clubpenalty=4000 \widowpenalty=4000 \sfcode`\.=1000\relax}

\newcount\@tempcntc
\def\@citex[#1]#2{\if@filesw\immediate\write\@auxout{\string\citation{#2}}\fi
  \@tempcnta\z@\@tempcntb\m@ne\def\@citea{}\@cite{\@for\@citeb:=#2\do
    {\@ifundefined
       {b@\@citeb}{\@citeo\@tempcntb\m@ne\@citea
        \def\@citea{,\penalty\@m\ }{\bf ?}\@warning
       {Citation `\@citeb' on page \thepage \space undefined}}%
    {\setbox\z@\hbox{\global\@tempcntc0\csname b@\@citeb\endcsname\relax}%
     \ifnum\@tempcntc=\z@ \@citeo\@tempcntb\m@ne
       \@citea\def\@citea{,\penalty\@m}
       \hbox{\csname b@\@citeb\endcsname}%
     \else
      \advance\@tempcntb\@ne
      \ifnum\@tempcntb=\@tempcntc
      \else\advance\@tempcntb\m@ne\@citeo
      \@tempcnta\@tempcntc\@tempcntb\@tempcntc\fi\fi}}\@citeo}{#1}}

\def\@citeo{\ifnum\@tempcnta>\@tempcntb\else\@citea
  \def\@citea{,\penalty\@m}%
  \ifnum\@tempcnta=\@tempcntb\the\@tempcnta\else
   {\advance\@tempcnta\@ne\ifnum\@tempcnta=\@tempcntb \else
\def\@citea{--}\fi
    \advance\@tempcnta\m@ne\the\@tempcnta\@citea\the\@tempcntb}\fi\fi}

\def\ps@crcplain{\let\@mkboth\@gobbletwo
     \def\@oddhead{\reset@font{\sl\rightmark}\hfil \rm\thepage}%
     \def\@evenhead{\reset@font\rm \thepage\hfil\sl\leftmark}%
     \let\@oddfoot\@empty
     \let\@evenfoot\@oddfoot}

\ps@crcplain                    

\newcommand{\AmS}{{\protect\the\textfont2
  A\kern-.1667em\lower.5ex\hbox{M}\kern-.125emS}}

\makeatother

\newcommand{\bee}{\begin{equation}}
\newcommand{\ene}{\end{equation}}

\title{Correspondence Principle in Quantum Gravity}

\author{Kirill A. Kazakov
\address{Moscow State University, Physics Faculty, Department of
Theoretical Physics, \\ 117234, Moscow, Russian Federation}
\address{Uppsala University, Physics Department,\\
SE-751 21, Uppsala, Sweden}
\thanks{E-mail: $Kirill.Kazakov@fysik.uu.se$}}

\begin{document}
\begin{abstract}
The problem of consistent formulation of the correspondence principle
in quantum gravity is considered. The usual approach based on the use
of the two-particle scattering amplitudes is shown to be in disagreement
with the classical result of General Relativity given by the Schwarzschild
solution. It is shown also that this approach fails to describe whatever
non-Newtonian interactions of macroscopic bodies.
An alternative interpretation of the correspondence principle is given
directly in terms of the effective action. Gauge independence of the
$\hbar^0$ part of the one-loop radiative corrections to the gravitational
form factors of the scalar particle is proved, justifying the interpretation
proposed. Application to the black holes is discussed.
\end{abstract}

\maketitle

Establishing the correspondence between classical and quantum
modes of description in the case of the theory of gravity displays
features quite different from those encountered in other theories
of fundamental interactions. Namely, the usual limiting procedure
of the transition from quantum to classical theory turns out to be
inapplicable in General Relativity. In electrodynamics, e.g., the
quantum potential, defined as the Fourier transform (with respect
to the momentum transfer from one particle to another) of a
suitably normalized two-particle scattering amplitude, takes the
form of the Coulomb law when the momentum transfer becomes small
as compared with the particles' masses, so the above mentioned
procedure is accomplished by tending the masses to infinity.
Contrary to this, in General Relativity, the radiative corrections
do not disappear in the limit: particle masses $\to \infty,$
because strength of the gravitational interaction of particles is
proportional to their masses. In fact, the relative value of the
radiative corrections to the classical Newton law, corresponding
to the logarithmical contribution to the gravitational form
factors of the scalar particle, is independent of the scalar
particle mass \cite{donoghue}. Yet, this contribution is
proportional to the Planck constant $\hbar.$ Newton's theory is
therefore correctly reproduced in the formal limit $\hbar \to 0.$

There is, however, a still more important aspect of the
correspondence between classical and quantum pictures of
gravitation. The Einstein theory, being essentially nonlinear,
demands quantum theory to reproduce not only the Newtonian form of
the particle interaction, but also all the nonlinear corrections
predicted by General Relativity. In this respect, the
above-mentioned peculiarity of the gravitational interaction,
namely, its proportionality to the masses of particles, is
manifested in the fact (also pointed out in Ref.~\cite{donoghue})
that, along with true quantum corrections (i.e., proportional to
the Planck constant $\hbar$), the loop contributions also contain
classical pieces (i.e., proportional to $\hbar^0$). Thus, an
important question arises as to relationship between these
classical loop contributions and the classical predictions of
General Relativity.

In analogous situation in the Yang-Mills theories, the correct
correspondence between classical and quantum theories is
guaranteed by the fact that all radiative corrections to the
particle form factors disappear in the limit: masses $\to \infty,$
thus providing the complete reduction of a given quantum picture
to the corresponding nonlinear classical solution. It is claimed
in Ref.~\cite{donoghue} that when collected in the course of the
construction of the gravitational potential from the
one-particle-reducible Feynman graphs, the aforesaid classical
contributions exactly reproduce the post-Newtonian terms given by
the expansion of the Schwarzschild metric in powers of $r_{g}/r,$
where $r_{g}$ is the gravitational radius.

However, as will be shown presently, the value of the potential
found in Ref.~\cite{donoghue} disagrees with the classical result
given by the Schwarzschild solution. The latter has the form
\begin{eqnarray}\label{sch1}&&
ds^2 = \left(1-\frac{r_g}{r}\right) c^2 d t^2 - \frac{d r^2}{1-\frac{r_g}{r}}
\nonumber\\&&
- r^2 (d\theta^2 + \sin^2\theta\ d\varphi^2),
\end{eqnarray}
\noindent
where $\theta, \varphi$ are the standard spherical angles, $r$ is the radial
coordinate, and $r_g = 2 G M/c^2$ is the gravitational radius of a
spherically-symmetric distribution of mass $M.$ The form of $d s^2$ given
by Eq.~(\ref{sch1}) is fixed by the requirements $g_{ti} = 0,$
$i = r, \theta, \varphi,$ $g_{\theta\theta} = - r^2.$ To compare the two
results, however, one has to transform Eq.~(\ref{sch1}) to the DeWitt gauge
\begin{eqnarray}\label{gauge}
\eta^{\mu\nu}\partial_{\mu} g_{\nu\alpha}
- \frac{1}{2}\eta^{\mu\nu}\partial_{\alpha} g_{\mu\nu} = 0,
\end{eqnarray}
\noindent
used in Ref.~\cite{donoghue}.

The $t, \theta, \varphi$-components of Eq.~(\ref{gauge}) are
already satisfied by solution (\ref{sch1}). To meet the remaining
condition, one substitutes $r = f(\tilde{r}).$ A simple
calculation shows that the function $f(\tilde{r})$ should expand
as
$$f(\tilde{r}) = \tilde{r}\left[1 + \frac{1}{2} \frac{r_g}{\tilde{r}}
+ \frac{1}{2} \left(\frac{r_g}{\tilde{r}}\right)^2 +
\cdot\cdot\cdot \right]$$ at large $\tilde{r}$ for
Eq.~(\ref{gauge}) to be satisfied to the order $r^2_g/r^2.$

In the new coordinates, $g_{00}$ component of the Schwarzschild solution
takes the following form (the tilde is omitted),
\begin{eqnarray}
g_{00} = 1 - \frac{r_g}{r} + \frac{r^2_g}{2 r^2}
+ O\left(\frac{r^3_g}{r^3}\right)
\nonumber
\end{eqnarray}
Taking the square root of $g_{00},$ we see that the classical
gravitational potential turns out to be equal to
\begin{eqnarray}\label{sch2}&&
\Phi^{\rm{c}}(r) =  - \frac{G M}{r} + \frac{G^2 M^2}{2 c^2 r^2}.
\end{eqnarray}
The post-Newtonian correction here is twice as small as that
obtained in \cite{donoghue}.

Thus, we arrive at the puzzling conclusion that in the classical limit,
the quantum theory of gravity, being based on the Bohr correspondence
principle, does not reproduce the Einstein theory it originates from.

However, before making such a conclusion, one should question
relevance of the definition of gravitational potential through the
scattering amplitudes. It may well turn out that the above
discrepancy arises because of the incorrect choice of the
quantum-field quantity to be traced back to the classical
potential.

In what follows, I give an interpretation of the correspondence
principle as applied to gravity, which naturally resolves the
above paradox \cite{kazakov1}.

\begin{figure}
\epsfbox[70, 70, 600, 800]{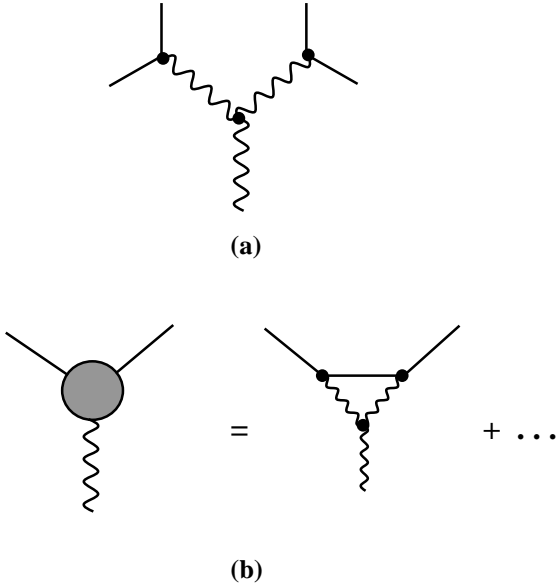} \vspace{-18cm}
\caption{Diagrams contributing to the first post-Newtonian
correction. (a) The tree diagram occurring because of the
nonlinearity of the Einstein equations. (b) The one-loop form
factor.  Wavy lines represent gravitons, solid lines scalar
particles.} \label{fig1}
\end{figure}

Let us note first of all that, from the formal point of view, the
correspondence between any quantum theory and its classical
original is most naturally established in terms of the effective
action rather than the S-matrix. This is because the effective
action (generating functional of the one-particle-irreducible
Green functions) just coincides with the initial classical action
in the tree approximation. In particular, the nonlinearity of
classical theory (resulting, e.g., in $r_g/r$ power corrections to
the Newton law in the General Relativity) is correctly reproduced
by the trees. In the case of quantum gravity, there are still
additional contributions of the order $\hbar^0,$ coming from loop
corrections. These are given by the gravitational form factors of
the particles, which serve as building blocks for the
gravitational potential. Instead of constructing the potential,
however, let us consider them in the framework of the effective
action method. From the point of view of this method, the
$\hbar^0$-parts of the particle form factors, together with the
proper quantum parts of order $\hbar,$ describe the radiative
corrections to the classical equations of motion of the
gravitational field. It is non-vanishing of these terms that
violates the usual Bohr correspondence. There is, however, one
essential difference between the $\hbar^0$ loop terms and the
nonlinear tree corrections. Consider, for instance, the first
post-Newtonian correction of the form
$${\rm const} \frac{G^2 M^2}{c^2 r^2}.$$ The "const" receives contributions
from the tree diagram pictured in Fig.~\ref{fig1}(a), as well as from the
one-loop form factor, Fig.~\ref{fig1}(b). If the gravitational field is
produced by only one particle of mass $M,$ then the two contributions
are of the same order of magnitude.

They are not, however, if the field is produced by a macroscopic
body consisting of a large number $N$ of particles with mass $m =
M/N.$ Being responsible for the nonlinearity of Einstein
equations, the tree diagram \ref{fig1}(a) is {\it bilinear} in the
energy-momentum tensor $T^{\mu\nu}$ of the particles, while the
loop diagram \ref{fig1}(b) is only {\it linear} (to be precise, it
has only two particle operators attached). Therefore, when
evaluated between the $N$-particle states, the former is
proportional to $(m \cdot N) \cdot (m\cdot N) = M^2,$ while the
latter, to $m^2 \cdot N = M^2/N.$ If, for instance, the solar
gravitational field is considered, the quantum correction is
suppressed by a factor of the order $m_{proton}/M_{\odot} \approx
10^{-57}.$
\vspace{-5cm}
\vspace{5cm}

This fact suggests the following interpretation of the
correspondence principle when applied to the case of gravity: {\it
the effective gravitational field produced by a macroscopic body
of mass $M$ consisting of $N$ particles turns into the
corresponding classical solution of the Einstein equations in the
limit $M \to \infty,$ $N \to \infty.$ }

It is clear from the above discussion that the use of the
effective action is essential for this interpretation: the quantum
potential defined with the help of the scattering amplitudes,
being of the form
\begin{eqnarray}&&
\Phi^{\rm{q}}(r) = - N \frac{G (M/N)}{r} + c_1 N \frac{G^2
(M/N)^2}{c^2 r^2}
\nonumber\\&&
+ c_2 N\frac{G^3 (M/N)^3}{c^4 r^3}
+ \cdot\cdot\cdot,
\nonumber
\end{eqnarray}
for a macroscopic ($N$-particle) body of mass $M,$ would fail to
reproduce whatever classical potential other than Newtonian, since
$$\Phi^{\rm{q}}(r) \to - \frac{G M}{r} ~~{\rm when} ~~N \to \infty.$$

Thus, the loop corrections of the order $\hbar^0$ are now
considered on an equal footing with the tree corrections, and
thereby are endowed with direct {\it physical meaning as
describing deviations of the spacetime metric from classical
solutions of the Einstein equations in the case of finite $N.$ }

Like any other argument trying to assign physical meaning to the
effective action, the above interpretation immediately runs up
against the problem of its gauge dependence. In spite of being
independent of the Planck constant, the $\hbar^0$ terms
originating from the loop diagrams are not gauge-independent {\it
a priori}. However, there is a strong evidence for that they {\it
are} gauge-independent nevertheless. Namely, as is demonstrated in
full detail in \cite{kazakov1}, the $\hbar^0$ terms of the
one-loop gravitational form factors of the scalar particle,
contributing to the first post-Newtonian correction to the metric,
turn out to be independent of the Feynman gauge parameter $\xi$
weighting the gauge condition. A very specific feature of this
gauge dependence cancellation must be emphasized: it holds only
for the $\hbar^0$ part of the form factors. For instance, the
logarithmical part of the form factors (which is of the order
$\hbar$) {\it is} gauge-dependent \cite{kazakov2}.

Actual value of the one-loop $\hbar^0$ contribution to the
effective metric can be found with the help of the explicit
expressions for the gravitational form factors of the scalar
particle, obtained in Ref.~\cite{donoghue}. In the momentum space,
\begin{eqnarray}&&\label{main1}
\delta g_{\mu\nu}(p) = - \frac{\pi^2 G^2 m^2}{c^2 \sqrt{-p^2}}
\nonumber\\&& \times \left(3 \eta_{\mu\nu} + \frac{q_{\mu}
q_{\nu}}{m^2} + \frac{7 p_{\mu} p_{\nu}}{p^2}\right),
\end{eqnarray}
\noindent where $q_{\mu}$ is the 4-momentum of the scalar particle
with mass $m.$

With the help of Eq.~(\ref{main1}), one can find the effective
gravitational potential of a spherically symmetric body with mass
$M,$ consisting of $N$ identical (scalar) particles with mass $m =
M/N,$ in the first post-Newtonian approximation. Namely, if the
gravitational interaction of the constituent particles is not too
strong (viz., if it can be represented as a series in powers of
$G$), then one has, in the coordinate space
\begin{eqnarray}&&\label{main}
\Phi^{\rm{eff}}(r) =  - \frac{G M}{r} + \frac{G^2 M^2}{2 c^2 r^2}
- \frac{G^2 M^2}{N c^2 r^2}.
\end{eqnarray}
\noindent
Indeed, the tree contribution to the effective field, represented in this
equation by the first two terms, always coincides with the corresponding
classical (Schwarzschild) solution. It is given, therefore,
by Eq.~(\ref{sch2}), all the effects of the particle interactions are
taken into account by identifying $M$ as the gravitational mass,
while account of the interaction in the loop correction (the third term in
Eq.~(\ref{main})), which is itself of the order $G^2,$ would give rise to
terms of the order $G^3,$ irrelevant to the present concern. Appearance of
the factor $1/N$ in the third term was explained earlier.

It was mentioned above that the loop contributions to the
post-Newtonian corrections are normally highly suppressed: their
relative value for the stars is of the order $10^{-56}-10^{-58}.$
This differs, however, in the case of objects consisting of
strongly interacting particle. In particular, in the limit of an
infinitely strong interaction, the object is to be considered as a
{\it particle}, i.e., one has to set $N = 1$ in Eq.~(\ref{main}).
The sign of the first post-Newtonian correction is then opposite
to that given by the classical General Relativity.

An example of objects of this type is probably supplied by the
black holes. Applicability of the above results to these objects
depends on whether they can be considered as elementary particles.
It seems that at least certain types of the black holes do behave
like normal elementary particles \cite{wilczek}. It should be
noted, however, that the very existence of the horizon is now
under question. The potential $\Phi^{{\rm eff}}(r)$ may well turn
out to be a regular function of $r$ when all the $\hbar^0$ loop
corrections are taken into account.

If the notion of elementary particle is indeed applicable to the
black holes, then the above results imply, in particular, that the
emission of the gravitational waves by the black hole binaries is
strongly affected by the quantum contributions already in the
first post-Newtonian approximation. The LIGO and VIRGO
gravitational wave detectors \cite{ligo}, which are now under
construction, will hopefully bring light into this issue.


\begin{thebibliography}{9}
\bibitem{donoghue}
~J.~F.~Donoghue, Phys. Rev. Lett. {\bf 72} (1994) 2996;
Phys. Rev. D{\bf 50} (1994) 3874;
{\it Perturbative Dynamics of Quantum General Relativity},
invited plenary talk at the "Eighth Marcel Grossmann Conference
on General Relativity", Jerusalem (1997).

\bibitem{kazakov1}
~K.~A.~Kazakov, Class. Quantum Grav. {\bf 18} (2001) 1039.

\bibitem{kazakov2}
~K.~A.~Kazakov, Phys. Rev. D{\bf 63} (2001) 044004.

\bibitem{wilczek}
~C.~F.~E.~Holzhey, F.~Wilczek, Nucl. Phys. {\bf B380} (1992) 447.

\bibitem{ligo}
Description  of the LIGO and VIRGO projects as well as information
about their current development is available at \\
http://www.ligo.caltech.edu
http://www.virgo.infn.it.

\end{thebibliography}
\end{document}